\documentclass[apj]{emulateapj}

\usepackage{apjfonts}
\usepackage{graphicx}
\usepackage{natbib}

\newcommand{\vt}{\mbox{\bf {T}}}
\newcommand{\vs}{\mbox{\bf {S}}}

\newcommand{\corr}{{\cal{\bf {R}}}}

\newcommand{\vn}{\mbox{N}}
\newcommand{\vnt}{\hat{\bf {n}}}

\newcommand{\cov}{\mbox{\bf {C}}}
\newcommand{\teff}{\tau_{\mbox{eff}}}
\newcommand{\ffi}{f_iB_{\ell m}^i}
\newcommand{\ffj}{f_jB_{\ell m}^j}

\newcommand{\gap}{ ^{~}_{>}}

\input epsf

\def\plotancho#1{\includegraphics[width=18cm]{#1}}

\begin{document}


\title{Tomography of the Reionization Epoch with Multifrequency CMB Observations}


\author{Carlos Hern\'andez--Monteagudo,\altaffilmark{1} Licia Verde,\altaffilmark{1} and Raul Jimenez\altaffilmark{1}}

\affil{}

\altaffiltext{1}{Department of Physics and Astronomy, University of Pennsylvania, 209 South 33rd St, Philadelphia, PA 19104; carloshm@astro.upenn.edu, lverde,raulj@physics.upenn.edu}


\begin{abstract} 
  
  We study the constraints that future multifrequency Cosmic Microwave
  Background (CMB) experiments will be able to set on the metal
  enrichment history of the Inter Galactic Medium at the epoch of
  reionisation. We forecast the signal to noise ratio for the
  detection of the signal introduced in the CMB by resonant scattering
  off metals at the end of the Dark Ages. We take into account
  systematics associated to inter-channel calibration, PSF
  reconstruction errors and innacurate foreground removal.  We develop
  an algorithm to optimally extract the signal generated by metals
  during reionisation and to remove accurately the contamination due
  to the thermal Sunyaev-Zel'dovich effect.
  Although demanding levels of foreground characterisation and control
    of systematics are required, they are very distinct from those
    encountered in HI-21cm studies and CMB polarization, and this fact
    encourages the study of resonant scattering off metals as an
    alternative way of conducting tomography of the reionisation
    epoch.  An ACT-like experiment with optimistic assumtions on
    systematic effects, and looking at {\em clean} regions of the sky,
    can detect changes of 3-12\% (95\% c.l.) of the OIII abundance
    (with respect its solar value) in the redshift range $z\in$
    [12,22], for reionization redshift $z_{\rm re}>10$. However, for
    $z_{\rm re} <10$, it can only set upper limits on NII abundance
    increments of $\sim$ 60\% its solar value in the redshift range
    $z\in$ [5.5,9], (95\% c.l.). These constraints assume that
    inter-channel calibration is accurate down to one part in
    $10^{4}$, which constitutes the most critical technical
    requirement of this method, but still achievable with current 
    technology.


\end{abstract}



\section{Introduction}

The Cosmic Microwave Background (CMB) radiation constitutes one of
today's most useful tools in cosmology. The properties of its
brightness temperature fluctuations carry information not only of the
young Universe in which they arise, but also of the evolving Universe
that CMB photons encounter in their way to the observer. First and
second generation CMB experiments like COBE \citep{COBE}, Tenerife
\citep{Tenerife}, Boomerang \citep{Boomerang}, Maxima \citep{Maxima} and WMAP \citep{WMAP,spergel06,
bennett03,hinshaw03,hinshaw06},
aim to characterize the properties of the {\em intrinsic} temperature
fluctuations, imprinted in the CMB during the epoch of Hydrogen
recombination at $z \sim 1,100$, \citep{zs}.  Third generation CMB
experiments, however, scan the small angular scales searching for {\em
  secondary} temperature fluctuations, introduced well after CMB
photons decouple from matter during recombination. Among all secondary
anisotropies, those induced by the thermal Sunyaev-Zel'dovich (tSZ)
effect (\citet{tSZ}, caused by the distortion induced by hot electrons
as they transfer energy to low energy CMB photons via Compton
scattering) and the kinematic Sunyaev-Zel'dovich effect (\citep{kSZ},
caused by a Doppler kick induced by moving electrons via Thomson scattering)
are of outmost importance. \\

Indeed, the fact that the tSZ effect is a {\em redshift-independent}
signal generated in hot dense electron clouds makes the tSZ a powerful
tool to measure the abundance of galaxy clusters at all redshifts,
which by itself can provide information about the presence and nature
of Dark Energy in our Universe, (e.g., \citet{DE01}). Third generation
experiments like South Pole Telescope (SPT, \citet{SPT}), 
 and the Atacama Cosmology Telescope (ACT, \citet{fowler}) will
scan the sky at millimeter frequencies in an attempt to characterise
the mass and redshift distribution of galaxy clusters in our Universe.\\

Here, we exploit the exquisite sensitivity level and angular
resolution of these forthcoming experiments to study the impact of
metals on CMB temperature fluctuations. Metals are first produced by
stars at the end of the Dark Ages and during reionization (for a
review see, e.g., \citet{BarkanaLoeb2001}), and, as shown in
\citet{basu}, they provide some effective optical depth to CMB photons
due to resonant transitions. \citet{basu} (hereafter BHMS) showed that
neutral species such as C, O and Si, together with ions like CII, OIII
or NII should leave their imprint on the CMB temperature anisotropy by
means of resonant scattering of CMB photons on some of their
transitions. They found that the leading consequence of such
scattering is a slight blurring of the original CMB anisotropies (by a
factor $\tau_{\nu_{rs}}$, with $\tau_{\nu_{rs}}$ the optical depth of
a resonant transition) in the temperature maps and
thus a contribution to the power spectrum proportional to the
primordial CMB power spectrum with constant of proportionality given
by $\tau_{\nu_{rs}}$.
Peculiar velocities of metals should introduce new anisotropies, which
however should only be visible in the very large angular scales. BHMS
demonstrated that, under {\em ideal} conditions, future multifrequency
CMB experiments should be able to set strong constraints on the
abundances of metals ($10^{-4} - 10^{-2}$ solar fraction) in the
redshift range $z\in [1-50]$, opening a new exploration window on the
end of the Dark Ages and the
reionization history of our Universe.\\

In this paper we build upon the work of BHMS paying particular
attention to realistic aspects such as cross-channel calibration,
point spread function (PSF) characterization and foreground
contamination, aiming to assess their real impact on the sensitivity
of future CMB experiments to metal enrichment history. We study what
level of experimental systematic and knowledge of contaminants is
required in order to achieve a given sensitivity threshold.  Our
approach is general, but we will focus on ACT's specifications
(e.g., instrumental noise level, angular resolution and sky coverage,
beam reconstruction etc.). Assuming a very accurate PSF
characterization and inter-channel calibration for this experiment, we
forecast ACT's sensitivity to the signal generated by metals and the
constraints on reionization history that can be inferred from it.
Current and future dust observations will help to correct for dust
emission to significantly low levels in already relatively dust-clean
sky areas. Also, because of the redshift-independent frequency
dependence of the non-relativistic tSZ, an accurate subtraction this
effect should be possible if more than two frequency channels are
available.
We find that the following specifications are required to extract the
signal from a realistic survey: cross channel calibration error at or
below the $10^{-4}$ level, PSF characterisation at $0.1-1$\% level,
efficient foreground removal. Foreground and PSF characterisation
requirements are realistically achievebable with an ACT-like
experiment. The required accuracy in cross-channel calibration can
certainly be achieved with Fourier Spectrometers,
\citep{fspec1,fspec2}. Such an experiment
will be able to set
strong constraints on the reionization history; in particular it
should detect the presence of newly formed OII at redshift $z\sim
14$ for metallicities of the inter-galactic medium above 3\% the solar
value, at the 95\% confidence level.
These constraints will provide a critical consistency check for
estimates of the optical depth to reionization and constraints on
reionization history obtained from CMB polarization measurements. In
addition these finding will provide a direct test of our understanding
of reionization and metal enrichment of the IGM.\\

Since the measurement of the metallicity abundance is sensitive to a
redshift window determined by the central frequency and the bandwidth
of the detector, multifrequency observations will provide a
tomographic reconstruction of the epoch of re-ionization. This should
motivate future experiments aiming to perform reionization tomography
via CMB multifrequency observations.  We offer guidance for these
future experiments and quantify their technical requirements.  In
Section 2 we briefly review how the signal we are trying to measure
arises and in Section 3 we build a realistic model of the total signal
to be measured by experiments. In Sections 4 and 5 we compare the
signal generated at reionization with contaminants, and suggest
different methods to extract it. In Section 6 we make predictions for
an ACT-like experiment: we build a realistic foreground model and
compute and discuss the limits that can be imposed on the reionisation
and enrichment history of the Universe.
We conclude in Sec. 7 discussing our results and prospects for future and
dedicated experiments.
%

\section{CMB resonant scattering on metals}
Following BHMS here we review how the signal of metals arises.
Let us assume that a CMB experiment is observing at a given frequency
$\nu_{obs}$ and that CMB photons interact with a species $X$ via a
resonant transition of resonant frequency $\nu_{rs}$. It is clear that
the CMB experiment may be sensitive to such interaction only in the
redshift range centered on $z_{rs} = \nu_{rs}/\nu_{obs} -1$, with its
width depending on the frequency response of the
detectors\footnote{The frequency bandwidth of the detectors is assumed
 to be much larger than the line thermal broadening}. The resonant
scattering of this species will generate an optical depth to CMB
photons which can be computed to be \citep{sobolev}:
\begin{equation}
\tau_{X} (z) = f_{rs} \frac{\pi e^2}{m_e c}\frac{\lambda_{rs} n_X(z)}{H(z)}.
\label{eq:tausob1}
\end{equation}
Here, $n_{X}$ is the number density of the $X$ species, $f_{rs}$ is the
absorption oscillator strength of the resonant transition, $\lambda_{rs}$
its wavelength, $H(z)$ is the Hubble function, $e$ and $m_e$ are
the electron charge and mass, respectively, $c$ is the speed of light.

The effect of resonant scattering on CMB anisotropies can be more easily 
seen when considering a single Fourier $k$-mode of the CMB temperature fluctuations.
Such mode can be expressed as an integral of various 
sources along the line of sight \citep{HS94}:
\[
\Delta_T (k,\eta_0,\mu) = \int_{0}^{\eta_0} e^{ik\mu(\eta-\eta_0)}
  \biggl[\Lambda(\eta )(\Delta_{T0}-i\mu v_b)
\]
\begin{equation}
\phantom{xxxxxxx} \dot{\phi} + \psi - ik\mu\psi \biggr].
\label{eq:dtk}
\end{equation}
In this equation, written for the conformal Newtonian gauge, $\eta$
denotes conformal time, $\Lambda (\eta) \equiv {\dot \tau}(\eta )
\exp{[-\tau (\eta)]}$ is the visibility function and $\tau (\eta )
\equiv \int_{\eta}^{\eta_0} d\eta ' {\dot \tau}(\eta ')$ is the
optical depth to CMB photons. For Thomson scattering, ${\dot \tau} =
a\sigma_T n_e$ with $a$ the scale factor, $n_e$ the electron number
density and $\sigma_T$ the Thomson cross section,  $\Delta_{T0}$
accounts for the intrinsic temperature fluctuations, and $v_b$ denotes
the Fourier mode for baryon peculiar velocities; $\phi$ and $\psi$
are the Fourier coefficients of the scalar perturbations of the
metric.

As the CMB is being observed at a given frequency $\nu_{obs}$, the
opacity due to the resonant transition can be approximated by
\begin{equation}
\dot{\tau}_{X} = \tau_X \; \delta_D (\eta - \eta_{rs} ),
\label{eq:opac_line}
\end{equation}
where $\eta_{rs}$ corresponds to $z_{rs}$, that is, the redshift at which observed
CMB photons could interact with the species $X$ via resonant scattering.
When plugging this opacity into eq.(\ref{eq:dtk}), one obtains that original
anisotropies have been 
blurred by a factor $e^{-\tau_X}$, and that new anisotropies
have been generated at the same redshift:
\begin{equation}
\Delta_T (k,\eta_0,\mu) = e^{-\tau_{X}} \Delta_{T_{orig}} (k,\eta_{rs},\mu)\; 
         + \; \Delta_{T_{new}} (k,\eta_{rs},\mu)
\label{eq:dtk2}
\end{equation}
In the limit of $\tau_X \ll 1$ (low metal abundance), we can retain
only linear terms in $\tau_X$. Thus the change in the anisotropies
($\Delta_T - \Delta_{T_{orig}}$) reduces to a {\em blurring} factor
($-\tau_X \Delta_{T_{orig}}$) plus a new term proportional to $v_b$,
which is important only in the large scales (see Appendix A of BHMS
for details). On the sphere, the blurring term is $\Delta (\delta T /
T_0) = - \tau_X (\delta T / T_0)_{CMB}$ and in multipole space,
$\Delta a_{\ell m} = - \tau_X a_{\ell m}^{CMB}$. Small scale
experiments will be only sensitive to the blurring term, which will
may be confused with other components present in the map. We address
this issue in the next 3 sections.

\section{Model of measured temperatures}

The fluctuations measured at each pixel location by a given channel
will be, in general, a sum of different components, namely intrinsic
CMB fluctuations, tSZ-induced temperature fluctuations, metal-induced
temperature fluctuations (which we aim to extract), galactic and
extragalactic contaminants and instrumental noise.  
Hence, the signal
received by  channel  $j$ observing a frequency $\nu_j$ is modeled as:\\
\begin{equation}
\vt_j = f_j \left[W_j *\left( \vt_{CMB} + \vt_{tSZ,j} + \sum_{M}\vt_{M,j}
   + \sum_{F} \vt_{F,j} \right)+ \tilde{ \vn}_j \right],
\label{eq:sm1}
\end{equation}
where $M$ stands for sum over {\it metals} and $F$ for sum over {\it
  foregrounds}, $\tilde{ \vn}$ denotes instrumental noise, and all $j$
indices refer to frequency $\nu_j$; $f_j$ is an overall factor
containing the calibration factor of the given channel, and $W_j$
represents its beam or point spread function (hereafter PSF). The symbol
$*$ denotes convolution. One can
rewrite the Fourier counterpart of the previous equation as:
\[
\tilde{ a}_{\ell m}^{j} =  \phantom{xxxxxxxxxxxxxxxxxxxxxxxxxxxx} 
\]
\begin{equation}
f_j  \biggl[ W_{lm}^j \left( a_{\ell m}^{CMB}
         + g_{tSZ,j}\;a_{\ell m}^{tSZ} + \sum_M g_{M,j}\; a_{\ell m}^{M} +
  \sum_{F} g_{F,j}\; a_{\ell m}^F               \right) 
+ \tilde{ \vn}_{lm}^{j}\biggr].
\label{eq:sm1F}
\end{equation}
This time we have split the peculiar frequency dependence of
temperature fluctuations of tSZ ($g_{tSZ,j}$), foregrounds ($g_{F,j}$)
and metals ($g_{M,j}$) from the multipole ($\ell$) dependence (i.e.
$g_{X,j}\equiv 1$ for a reference channel).  However to keep track
explicitly of the ``real life effects'', we will work with the
$a_{\ell m}^j$'s, defined as:
\[
a_{\ell m}^j \equiv \frac{ \tilde{ a}_{\ell m}^{j} }{(W_{\ell m}^j)^{model}} =
 \phantom{xxxxxxxxxxxxxxxxxxxxxxxxxxx}
\]
\begin{equation}
 f_j  \left[ B_{\ell m}^j  \left( a_{\ell m}^{CMB}
         + g_{tSZ,j}\;a_{\ell m}^{tSZ} + \sum_M g_{M,j}\; a_{\ell m}^{M} +
  \sum_{F} g_{F,j}\; a_{\ell m}^F \right) + \vn_{\ell m}^{j}\right].
\label{eq:almr1}
\end{equation}

In this equation, $B_{lm}^j\equiv  W_{lm}^j/ (W_{lm}^j)^{model} =
1 + \epsilon_{\ell m}^j$
expresses the departure of the {\em real}
PSF from the {\em model} PSF, (i.e. if the PSF is perfectly determined,
$\epsilon_{\ell m}^j = 0$ and $B_{lm}^j = 1$). 
The noise has been redefined as $\vn_{lm}^{j}
\equiv \tilde{ \vn}_{lm}^{j} / (W_{lm}^j)^{model}$. Finally, if we set
some frequency channel $i$ to be the {\it reference} channel,  we
can write the calibration factor $f_j$ as 
\begin{equation}
f_j = 1 + \eta_j,
\label{eq:calf1}
\end{equation}
where $\eta_j$ denotes the error in the relative calibration with
respect to channel $k$. As working hypotheses, we shall assume that
both calibration and PSF characterization are {\em very} accurate,
i.e., $\eta^j \ll 1$ and $|\epsilon^j_{\ell m}| \ll 1$ .\\

The signal induced by metals can be rewritten as
\begin{equation}
\sum_M g_{M,j}\; a_{\ell m}^{M} = -\tau_{\mbox{eff}}^j \;a^{CMB}_{\ell m},
\label{eq:almm}
\end{equation}
where the frequency dependence of the effective optical depth measures the
amount of metals present in the redshift range probed by the observing
frequency and the frequency of the resonant transition ($rs$), $z_{rs}
= \nu_{rs}/\nu_{obs} - 1$. The lowest frequency channel will
correspond to the highest redshift range, for which we will assume
that the metal abundance is lowest, and therefore will be regarded as
{\em reference} channel.  We will attempt to measure {\em increments}
of metal abundances with respect to such redshift by studying the
difference map and the difference power spectra between different
channels.

\section{Power spectrum of difference map versus difference of power spectra}
\subsection{The difference map}
The difference map between channels $j$ and $i$ in multipole space is,
$$
\delta a^{ji}_{\ell m} \equiv a^j_{\ell m} - a^i_{\ell m} = 
 \phantom{xxxxxxxxxxxxxxxxxxxxx}
$$
$$
a^{CMB}_{\ell m} 
 \biggl[\ffj - \ffi - (\ffj\;\teff^j - \ffi\;\teff^i)\biggr] \; + 
$$
$$
 a_{\ell m}^{tSZ} \biggl[\ffj\; g_{tSZ,j} - \ffi\;g_{tSZ,i}\biggr] +
\sum_F a_{\ell m}^F \biggl[\ffj\; g_{F,j} - \ffi\;g_{F,i}\biggr] 
$$
\begin{equation}
\phantom{xxxxxxxxxxx} +  f_j\vn_{\ell m}^j - f_i \vn_{\ell m}^i.
\label{eq:deltaalm}
\end{equation}
Since we assume accurate calibration and PSF characterization
 $\ffj - \ffi \simeq 2(\eta_j - \eta_i +
 \epsilon^j_{\ell m}-\epsilon^i_{\ell m}) \equiv 2(\delta f_{ji} + \delta
 \epsilon_{\ell m}^{ji})$ and $\ffj\;g_{X,j} - \ffi\;g_{X,i} \simeq
 g_{X,j} - g_{X,i}$ for any frequency-dependent component $X$. Hence,
 eq.~(\ref{eq:deltaalm}) can be re-written as
$$
\delta a^{ji}_{\ell m} \simeq a^{CMB}_{\ell m} 
 \biggl[2(\delta f_{ji} + \delta\epsilon_{\ell m}^{ji}) - \delta
 \teff^{ji}\biggr] \; + a_{\ell m}^{tSZ} \biggl[g_{tSZ,j} -
 g_{tSZ,i}\biggr] +
$$
\begin{equation}
\sum_F a_{\ell m}^F \biggl[g_{F,j} - g_{F,i}\biggr] + 
   f_j\vn_{\ell m}^j - f_i \vn_{\ell m}^i,
\label{eq:deltaalm2}
\end{equation}
with $\delta \teff^{ji} \equiv \teff^j - \teff^i$.\\

This equation already shows the main limiting factor in our procedure:
the accuracy of the cross-channel calibration. If $\delta f_{ji}$ is
negligible, then it should be possible to filter the $\delta
a^{ji}_{\ell m}$ to recover $\delta \teff^{ji}$ and discriminate it
from the remaining signals, which should show, a priori, a {\em
  different} spatial pattern\footnote{We are assuming that errors in
  the PSF characterization are {\em uncorrelated} with the intrinsic
  CMB field, making the field $\delta\epsilon_{\ell
    m}^{ji}a^{CMB}_{\ell m}$ spatially independent of $a^{CMB}_{\ell
    m}$}. However, if $\delta f_{ji}$ is non zero, then it will give
rise to a spurious signal of identical spatial properties to the
signal we are after ($-\delta \teff^{ji}\;a^{CMB}_{\ell m}$), and
hence indiscernible from it. Therefore, the calibration uncertainties
set an unavoidable limit to the sensitivity of a CMB experiment that
aims to measure the signal
induced by scattering on metals during reionization.\\

\subsection{Difference of power spectra versus power spectrum of difference map.}

We next investigate the possibility of detecting the metal-induced
signal in the power spectrum. There are two relevant quantities in
this context: the difference of the power spectra computed at two
different frequencies, $\delta C_l^{ji} \equiv C_l[\vt_j] -
C_l[\vt_i]$ and the power spectrum of the difference map, $C_l [\vt_j
- \vt_i]$. The latter consists on an average of $|\delta a_{\ell
m}^{ji}|^2$, which, according to eq.~(\ref{eq:deltaalm2}), reads
$$
|\delta a_{\ell m}^{ji}|^2 = |a^{CMB}_{\ell m}|^2 
   \bigg[ |\delta f_{ji} + \delta\epsilon_{\ell m}^{ji}|^2 + 
  (\delta \teff^{ji})^2 - 2\delta \teff^{ji}{\cal R}[\delta f_{ji} +  
  \delta\epsilon_{\ell m}^{ji}]\biggr] \;
$$
$$ 
+ \;
 |a_{\ell m}^{tSZ}|^2 (\Delta g^{ji}_{tSZ})^2 \; + \;
\sum_F |a_{\ell m}^{F}|^2 (\Delta g^{ji}_{F,F})^2 \; 
$$
$$
+ \;
   2\sum_{F_u > F_v} {\cal R}[a_{\ell m}^{F_u} (a_{\ell m}^{F_v})^*]
                             \Delta g^{ji}_{F_u}\Delta g^{ji}_{F_v}  + 
    {\cal O}[\delta \vn_{\ell m}^2] \; 
$$
$$
+\; 2{\cal R}[a_{\ell m}^{CMB}(\delta f_{ji} + \delta\epsilon_{\ell m}^{ji}) 
                                   (a_{\ell m}^{tSZ})^*] \Delta g^{ji}_{tSZ} 
$$
$$
-
\; 2{\cal R}[a_{\ell m}^{CMB}(a_{\ell m}^{tSZ})^*] \delta 
                \teff^{ji}\Delta g^{ji}_{tSZ} \;
$$
$$
+\; 2 \sum_F {\cal R}[a_{\ell m}^{CMB}(\delta f_{ji} + \delta\epsilon_{\ell m}^{ji})
                       (a_{\ell m}^{F})^*]  \Delta g^{ji}_{F} \;
$$
$$ 
- \;
2 \sum_F {\cal R}[a_{\ell m}^{CMB}(a_{\ell m}^{F})^*] \delta 
                \teff^{ji}\Delta g^{ji}_{F} 
$$
\begin{equation}
+\; 2 \sum_F {\cal R}[a_{\ell m}^{tSZ}(a_{\ell m}^{F})^*]  \Delta g^{ji}_{tSZ} 
                          \Delta g^{ji}_{F} \; + \; 
                           {\cal O}[\mbox{cross terms }\vn_{\ell m}].
\label{eq:deltaalmsq}
\end{equation}

For third generation CMB experiments, instrumental noise will not be a
 limiting factor, and therefore here we have neglected the cross terms
 where they appear. $\Delta g^{ji}_X \equiv g_{X,j} - g_{X,i}$ for any
 frequency dependent component $X$, and ${\cal R}[Y]$ denotes the real
 part of $Y$; $F_u$ and $F_v$ denote different foreground components.
 This equation simplifies considerably under the assumption that
 $\delta f_{ji} + \delta\epsilon_{\ell m}^{ji}$ and $\delta
 \teff^{ji}$ are much smaller than the typical amplitudes of tSZ and
 foregrounds. In this case, the average ($\sum_m/(2\ell +1)$) of
 eq. ({\ref{eq:deltaalmsq}) becomes:
$$
\langle |\delta a_{\ell m}^{ji}|^2 \rangle \approx 
    \langle|a_{\ell m}^{tSZ}|^2\rangle (\Delta g^{ji}_{tSZ})^2 \; + \;
\sum_F \langle |a_{\ell m}^{F}|^2\rangle (\Delta g^{ji}_{F,F})^2 \;
$$
\begin{equation}
 + \;
   2\sum_{F_u > F_v} {\cal R}\langle[a_{\ell m}^{F_u} (a_{\ell m}^{F_v})^*]\rangle
                             \Delta g^{ji}_{F_u}\Delta g^{ji}_{F_v}
\label{eq:deltaalmsq2}
\end{equation}
The last term in eq.  ({\ref{eq:deltaalmsq}) has been neglected
because the correlation between tSZ and foregrounds becomes negligible
when massive cluster are excised from the maps (see
sec. \ref{sec:ACT}).  That is, the power spectrum of the difference
map provides a template for the dominant foregrounds in the case where
the metal-induced signal is subdominant in $\vt_j - \vt_i$. We will
show in Section (\ref{sec:tsz}) that the spatial properties of
foregrounds inferred from this statistic can be used to optimize the
search for signatures of reionization.\\

We next compute the difference of the power spectra computed at the
channels considered, $j$ and $i$. Noting that 
\begin{equation}
\delta C_{\ell}^{ji} \equiv C_{\ell}[\vt_j] - C_{\ell}[\vt_i] = 
  \frac{1}{2\ell+1}\sum_m \biggl(|a_{\ell m}^j|^2 -|a_{\ell m}^i|^2\biggr),
\label{eq:dcldef}
\end{equation}
a straightforward calculation for $|a_{\ell m}^j|^2 -|a_{\ell m}^i|^2$ yields
$$
|a_{\ell m}^j|^2 -|a_{\ell m}^i|^2 \simeq 2\;|a_{\ell m}^{CMB}|^2 \biggl(
{\cal R}[\delta f_{ji} + \delta\epsilon_{\ell m}^{ji}] \; - \;
\delta \teff^{ji} \biggr) \;  
$$
$$
+ \; |a_{\ell m}^{tSZ}|^2 \Delta[(g_{tSZ}^{ji})^2] \; + \;
\sum_F |a_{\ell m}^{F}|^2 \Delta[(g_{F}^{ji})^2] 
$$
$$
\phantom{xxxxxxxxxx}\; + \;
2\sum_{F_u > F_v} {\cal R}[a_{\ell m}^{F_u} (a_{\ell m}^{F_v})^*] 
            \Delta[g_{F_u}^{ji}g_{F_v}^{ji}] \; 
$$
$$
+ \; |\vn_j|^2 - |\vn_i|^2 \; + \; 
 2{\cal R}[a_{\ell m}^{CMB} (a_{\ell m}^{tSZ})^*] \Delta[g_{tSZ}^{ji}] 
$$
$$
 \; + \;
2\sum_F {\cal R}[a_{\ell m}^{CMB} (a_{\ell m}^{F})^*] \Delta[g_{F}^{ji}] \; + \;
2\sum_F {\cal R}[a_{\ell m}^{tSZ} (a_{\ell m}^{F})^*] 
                 \Delta[g_{CMB}^{ji}g_{F}^{ji}] \;
$$
\begin{equation}
 + \;
{\cal O}[(\delta \teff^{ji})^2] \;+ \; {\cal O}[\mbox{cross terms }
                                 \delta\teff^{ji}] 
\label{eq:dcl2}
\end{equation}

The first term in the RHS of this equation reminds us of the
importance of the calibration, since no signal will be measured if
$\delta f^{ji} \sim \delta \teff^{ji}$. The requirements on the
amplitude of the PSF error $\delta \epsilon^{ji}$ are less strict than
those of $\delta f^{ji}$, since its real part can flip sign and
eventually will average to zero after combining different modes. Note
that our signal is proportional to $\delta \teff^{ji}$, and {\em not}
to $(\delta \teff^{ji})^2$, \citep{hms}.  Since our analysis is
restricted to small ($l>200$) scales, the correlation between the
Integrated Sachs-Wolfe (ISW) component and the tSZ and extragalactic
foregrounds can be neglected, because these signals are only relevant
at large angles. Indeed, the tSZ and ISW correlation shows small ($\sim$ 3--1
$\mu$K) amplitudes at $l>300$, \citep{hms}.  The correlation between
the tSZ and other foregrounds (e.g. radio galaxies) is also negligible
once bright galaxy clusters are excised out.  If we further neglect
noise (as done in eq. 12 above), then we are left with

$$
\langle |a_{\ell m}^j|^2 -|a_{\ell m}^i|^2 \rangle 
\approx 2\; \langle |a_{\ell m}^{CMB}|^2 \rangle \biggl(
{\cal R}[\langle \delta f_{ji} + \delta\epsilon_{\ell m}^{ji}] \rangle \; - \;
\delta \teff^{ji} \biggr) \; + \; 
$$
$$
\langle |a_{\ell m}^{tSZ}|^2 \rangle \Delta[(g_{tSZ}^{ji})^2] \; + \;
\sum_F \langle |a_{\ell m}^{F}|^2 \rangle \Delta[(g_{F}^{ji})^2] \;
$$
\begin{equation}
 + \;
2\sum_{F_u > F_v} \langle {\cal R}[a_{\ell m}^{F_u} (a_{\ell m}^{F_v})^*] \rangle 
            \Delta[g_{F_u}^{ji}g_{F_v}^{ji}] 
\label{eq:dcl3}
\end{equation}

If tSZ is the only significant contaminant in the maps, then the
$\ell$ shape of its power spectrum can be obtained from the difference
map power spectrum, eq.(\ref{eq:deltaalm2}), and removed from
$|a_{\ell m}^j|^2 -|a_{\ell m}^i|^2$. Note that this is only possible
because the spectral dependence of the tSZ is known to high accuracy
and bright clusters have been excised.  In more adverse scenarios,
eq.(\ref{eq:deltaalm2}) should always provide a rough estimate of the
amplitude and shape of the foreground power spectrum present in
$|a_{\ell m}^j|^2 -|a_{\ell m}^i|^2$. Note that the signal we are
searching for has a definite well know $\ell$-pattern, identical to
that of $C_{\ell}^{CMB}$.\\

\section{Applying a template fit}
\label{sec:fit}

Since we know what the spatial pattern of the signal we are trying to
unveil is; $ \vt_{ji} [\vnt ] = -\delta \teff^{ji} \vt^{CMB}[\vnt ]$,
with $\vnt$ denoting a position on the sky, and we are aiming to
measure one single parameter ($\delta \teff^{ji}$), it is justified to
write the following model for the measured difference map:

\begin{equation}
(\vt_j - \vt_i)[\vnt ] = \vn [\vnt ] -\delta \teff^{ji} \vt^{CMB}[\vnt ],
\label{eq:mod1}
\end{equation}
where $\vn$ contains all the contaminating signals.

If we assume that the signal $\vn $ is gaussianly
distributed\footnote{If this is not the case, our matched filter of
eq.(\ref{eq:tau1}) still provides an unbiased estimate of $\delta
\teff^{ji}$}, then an optimal estimator for $\delta \teff^{ji}$ is
given by \citep{alphamethod}:
\begin{equation}
\mbox{E}[\delta \teff^{ji}] = 
 - \frac{(\vt_j - \vt_i)^t\cov^{-1}\vt^{CMB}}{(\vt^{CMB})^t\cov^{-1}\vt^{CMB}},
\label{eq:tau1}
\end{equation}
with $\cov$ the covariance matrix of ${\bf N}$ (i.e. all the spurious
signals regarded as noise), which is simply the Legendre transform of
eq.(\ref{eq:deltaalm2}). The formal error associated to
$\mbox{E}[\delta \teff^{ji}]$ is then given by
\begin{equation}
\sigma_{\delta \teff^{ji}}^2 =  \frac{1}{(\vt^{CMB})^t\cov^{-1}\vt^{CMB}}.
\label{eq:sgtau1}
\end{equation}

One must keep in mind, however, that the inversion of $\cov$ might not
be trivial, since its dimensions are $N_{px} \times N_{px}$, with
$N_{px}$ the number of pixels of the map to be analyzed.  If $N_{px}$
is greater than a few thousands, the inversion of the covariance
matrix poses numerical problems. We adopt the approach presented in
\citet{isw1} of dividing the map into different equal-sized patches
for which the covariance matrix inversion is doable, and then
combining the estimates of $\delta \teff^{ji}$ obtained from each
patch.  Therefore, for each patch $\beta$, we obtain an estimate of
the increment in optical depths $\delta \teff^{ji,\beta}$ and an
associated error $\sigma_{\delta \teff^{ji,\beta}}$ given by
\begin{equation}
\mbox{E}[\delta \teff^{ji,\beta}] = -
 \frac{ (\vt_{j,\beta}-\vt_{i,\beta})^t 
               \bigl({\cov^{\beta,\beta}}\bigr)^{-1} 
                \vt^{CMB}_{\beta}} 
        {(\vt^{CMB}_{\beta})^t\bigl({\cov^{\beta,\beta}}\bigr)^{-1} 
                        \vt^{CMB}_{\beta} }
\label{eq:tau2}
\end{equation}
and
\begin{equation}
\;\; \sigma_{ \delta \teff^{ji,\beta} }^2 = 
             {\frac{1}{(\vt^{CMB}_{\beta})^t\bigl({\cov^{\beta,\beta}}
                                        \bigr)^{-1} \vt^{CMB}_{\beta}}}.  
\label{eq:tau2b}
\end{equation} 

We combine the $\delta \teff^{ji,\beta}$ estimates into a variance-weighted
single estimate
\begin{equation}
\mbox{E}[\delta \teff^{ji}] =  \frac{\sum_{\beta} 
       \mbox{E}[\delta \teff^{ji,\beta}]/(\sigma_{\delta \teff^{ji,\beta}})^2}
                  {\sum_{\beta} 1/(\sigma_{\delta \teff^{ji},\beta})^2},
\label{eq:tau3} 
\end{equation}
whose uncertainty takes into account the fact that estimates of 
$\delta \teff^{ji,\beta}$ for {\em different} patches may not be independent:
\[
\;\; \sigma_{\delta \teff^{ji}}^2 = 
     \frac{1}{\sum_{\beta} 1/(\sigma_{\delta \teff^{ji,\beta}})^2}
 \biggl[ 1 + 2\;\times
\]
\begin{equation}
\frac{\sum_{\beta_1<\beta_2}     (\vt^{CMB}_{\beta_1})^t 
                \bigl( {\cov}^{\beta_1,\beta_1} \bigr)^{-1} \; 
               {\cov}^{\beta_1,\beta_2} \;
              \bigl({\cov}^{\beta_2,\beta_2}\bigr)^{-1} \vt^{CMB}_{\beta_2}}
              { \sum_{\beta} 1/(\sigma_{\delta \teff^{ji,\beta}})^2 } \biggr].
\label{eq:sg_tau3}
\end{equation}
The second term in brackets vanishes if different patches are uncorrelated,
i.e., ${\cov}^{\beta_1,\beta_2}=0$ for $\beta_1 \ne \beta_2$.

\begin{figure*}
\centering
\plotancho{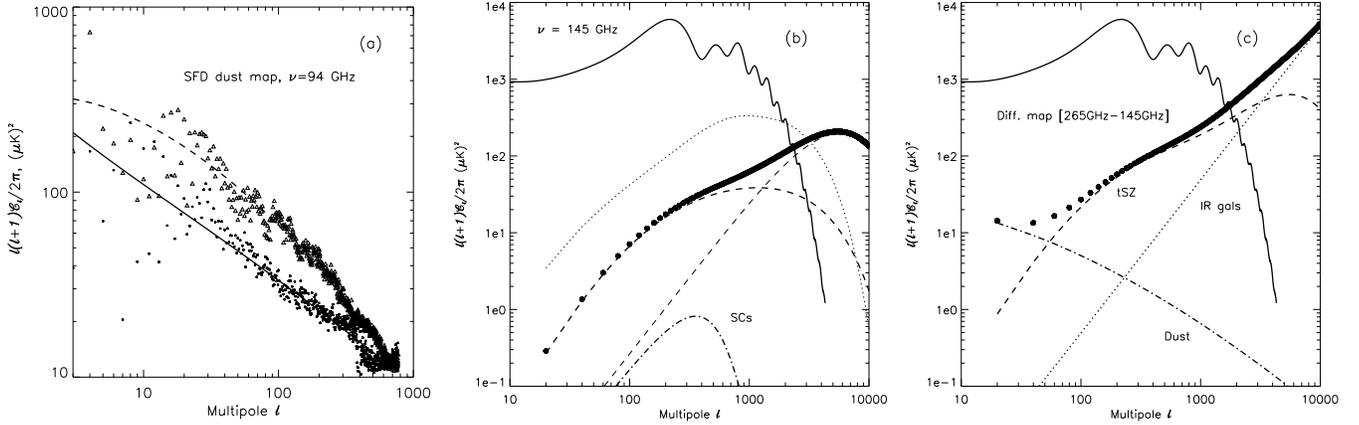}
\caption[fig:fig1]{{\it (a)} Dust temperature power spectra as
  computed from the dust maps available at LAMBDA. Kp2 (empty
  triangles) and Kp0 (filled circles) masks have been applied, and the
  analytical fits used in our foreground forecast model. {\it (b)} tSZ
  signal present at 145 GHz. Once  galaxy clusters above ACT's
  4-$\sigma$ threshold have been removed (dotted line), only galaxy
  groups and diffuse gas in superclusters and filaments remain. The
  total tSZ residual (filled circles) is dominated by the galaxy group
  contributions, both from the Poissonian (thin dashed line) and the
  correlation (thick dashed line). The contribution of diffuse gas is
  negligible (dot-dashed line). {\it (c)} Power spectrum of the 265
  GHz -- 145 GHz difference map assuming no metal signal (filled
  circles), and its three components: tSZ (dashed line), IR sources
  (dotted line) and galactic dust (dot-dashed line).}
\label{fig:fig1}
\end{figure*}

\section{Prospects for ACT}
\label{sec:ACT}
The Atacama Cosmology Telescope (ACT, \citet{fowler}) will scan around
200 clean square degrees in the southern equatorial hemisphere in
three frequency bands: 145 GHz, 220 GHz and 265 GHz. Its angular
resolution and sensitivity levels (FWHM $\simeq 1'$, with $\sim 2
\mu$K of instrumental noise per pixel) have been optimized to study
the tSZ effect in clusters of galaxies. Using these technical
specifications we make predictions on the sensitivity of this
experiment to the presence of metals at the end of the Dark Ages. We
find that, beyond the strict requirements of cross-channel calibration
and PSF characterization, the main limiting factor will be the
presence of un-subtracted tSZ signal together with galactic dust
emission.  In what follows we will consider two cases, a pessimistic
and an optimistic case, chosen to span the reasonable range of
possibilities.

\subsection{Foreground model}
\label{sec:foreg}

At the frequencies of ACT, we consider three main foreground
contaminants: galactic dust, tSZ and infrared (IR) galaxies.
Radio galaxies should contribute with sub-dominant contamination at
these relatively high frequencies, and their spatial properties should
be close to those of IR galaxies. \\

{\it Dust}: We build a model for dust emission based on the dust
antenna temperature map provided by the WMAP team \citep{wmapforeg}
and available at the LAMBDA site\footnote{LAMBDA URL site: {\it
http://lambda.gsfc.nasa.gov}}; this model was produced using the data
of \citet{fds}. After converting it into thermodynamic temperature, we
have computed its power spectrum outside foreground masks Kp0 and Kp2,
(open triangles for Kp2, filled circles for Kp0 in
Fig.\ref{fig:fig1}a), and fit it with power laws. This should provide
a fair estimate of dust emission at ACT's frequencies. In reality
however, dust cleaning of the actual data would need to be done
directly on the maps.  The maps have been scaled to ACT's frequencies
by using the intensity spectral index given in \citet{wmapforeg}
($\alpha = 2.2 \pm 0.1$). Note that this is a conservative choice:
this spectral index overpredicts the dust emission if compared to the
shallower index ($\alpha \simeq 1.7$) measured by \citet{archeops}.
We restrict our analyses to the {\em cleanest} third portion of the
available sky in ACT's strip, for which the variance of dust-induced
anisotropies is $\sim 4$ \% of the variance outside the Kp0
mask. Therefore, for our clean pixel subset we scale the dust power
spectrum amplitude accordingly.  We further assume that, after
removing the existing dust model, the residuals vary from 10\%
(optimistic case) to 30\% (pessimistic case) of the un-subtracted
emission (in thermodynamic temperature). The dot-dashed line in
Fig.(\ref{fig:fig1}c) shows the dust residuals in the power spectrum
of the 265 GHz -- 145 GHz difference map in the pessimistic case.\\

{\it tSZ signal:} The first step consists in removing all clearly
detected sources. In practice this means that all pixels showing tSZ
signals above the $\sim$ 4-$\sigma$ level should be excised; this
excludes practically all clusters above $\sim 10^{14} M_{\odot}$ (see
dotted line in Fig.\ref{fig:fig1}b).  We are then left with the tSZ
signal coming from smaller galaxy groups and the diffuse tSZ from a
warm intergalactic gaseous component. When computing the power spectra
of clusters and galaxy groups we consider both the Poisson term
($C_{\ell}^P$, \citet{atriojan}) and the correlation term
($C_{\ell}^C$, \citet{kk99})\footnote{The re-derivation of these terms
in a line of sight approach \citep{ksz} yielded expressions of smaller
amplitude. Since the tSZ is regarded as a foreground, we
conservatively use the equations given above.}.  These terms are given
by
\begin{equation}
C_l^P = \int dz \frac{dV(z)}{dz}\; \int_{M_{min}}^{M_{max}}dM \frac{dn}{dM}
          [M,z]\; |y_l(M,z)|^2
\label{eq:clp}
\end{equation}
and 
\[
C_l^C = \int dz \frac{dV(z)}{dz}\; P_m(k=\frac{l}{r(z)},z) \times
\]
\begin{equation}
\phantom{xxxx}
\biggl[ \int_{M_{min}}^{M_{max}}dM \frac{dn}{dM}[M,z] b(M,z) y_l(M,z) \biggr]^2
\label{eq:clc}
\end{equation}
For galaxy groups we chose $M_{min}=5\times 10^{12} M_{\odot}$ and
$M_{max} = 5\times 10^{13} M_{\odot}$; $y_l(M,z)$ is the Fourier transform of
the tSZ angular profile of a cluster,
\begin{equation}
y(\vnt ) = \int dr \sigma_T n_e \frac{k_B T_e}{m_e c^2},
\label{eq:y}
\end{equation}
where $\sigma_T$ is the Thomson cross-section, $k_B$ the Boltzmann
constant and $T_e$, $m_e$ and $n_e$ the electron temperature, mass and
density, respectively; $b(M,z)$ is the bias factor of \citet{mow96}
relating the halo power spectrum to the underlying matter power
spectra $P_m(k,z)$. Since we are interested in scales larger than the cluster,
we took a gaussian profile for the gas distribution within these sources.
Finally, we adopt a Press-Schechter mass
function to characterise the mass and redshift distribution of
collapsed objects.  Since galaxy groups are smaller than galaxy
clusters, the Poissonian term of the groups peaks at smaller angular
scales than for clusters (see thin dashed line in Fig.
(\ref{fig:fig1}b)). However, the distribution of galaxy groups is
highly correlated, and this correlation dominates at large angles
(thick dashed line in Fig. (\ref{fig:fig1}b)). These two terms are
responsible for virtually all the tSZ residual signal present in the
maps after subtracting the brightest galaxy clusters (solid circles).
We associated diffuse gas into massive ($M > 10^{15} M_{\odot}$)
structures that have turned around recently and hence still keep their
linear size (superclusters).  These structures are assumed to
have an average overdensity of $\delta_{SC} \sim 20$ and a gas
temperature of 0.2 KeV. These parameters should over-predict the
diffuse tSZ signal and hence provide an upper limit; still this is
negligible compared to the emission generated in galaxy groups. In
Fig.(\ref{fig:fig1}b) the dot-dashed line gives the amplitude of the
diffuse tSZ, which barely reaches the
1 $\mu$K level at sub-degree scales.\\

{\it Infrared galaxies}: Infrared galaxies present in ACT fields will
be observed by BLAST\footnote{BLAST URL site: {\it
http://chile1.physics.upenn.edu/blastpublic/index.shtml}}.  Therefore,
the major source of uncertainty should not be the amount of galaxies
present within each ACT PSF, but how the measured flux at BLAST
frequencies should be extrapolated at ACT frequencies. BLAST has three
channels at 600 GHz, 860 GHz and 1200 GHz, and depending on the
galaxy's redshift these frequencies will probe the dust emission
spectrum peak (high redshift) or the long wavelength tail (low
redshift).
To estimate flux of these sources at ACT lower frequencies, we need an
estimate specific intensity spectral index $\alpha$ in the long
wavelength tail, (i.e.  $I_{\nu} \propto \nu^{\alpha}$ but $\alpha$ is
expected to vary as a function of frequency).

We thus build the following model for the infrared galaxies foreground
signal and its residuals.  We use the results of \citet{knox04},
which, based upon SCUBA\footnote{SCUBA URL site: {\it
    http://www.ifa.hawaii.edu/users/cowie/scuba/scuba.html}}
observations, provides a spectral index of $\alpha_{SCUBA} = 2.6 \pm
0.3$ for the IR intensity over the range frequency 150-350 GHz.  We
use $\alpha_{SCUBA}$ to extrapolate IR source amplitudes from SCUBA to
ACT frequencies. We assign BLAST sources an effective spectral index
of $\alpha_{BLAST} \approx 2$ (shallower than that of SCUBA) to
convert intensities between BLAST and ACT frequencies. This fixes our
IR source model.
We then introduce an error ($\delta \alpha_{BLAST}$) in the spectral
index of 5\% (optimistic case) and 20\% (pessimistic case) of the
model value of $\alpha_{BLAST}$, and regard the subsequent error in
extrapolated intensity as and estimate of residuals IR source
subtraction.  We neglect any type of spatial correlation of IR sources
and hence the IR source residuals show a Poisson spectrum (see dotted
line in Fig.~\ref{fig:fig1}c, which corresponds to $\delta
\alpha_{BLAST} / \alpha_{BLAST} = 0.2$). \\


We see that infrared galaxies contaminate ACT maps at small scales,
particularly at 265 GHz. This level of contamination can be lowered by
excising out the brightest galaxies, but here we prefer to compute the
forecasts without masking out more area.  After all, at the angular
scales at which the intrinsic CMB is dominant, the IR galaxy
contribution is still below the tSZ residuals. Only at large angular
scales the contribution from galactic dust is
relevant. Fig.~\ref{fig:fig1}c shows our expectation for the power
spectrum of the 265 GHz -- 145 GHz difference map (filled circles),
assuming that there is no metal-induced signal, together with its
three components for our pessimistic case. This is the most
contaminated scenario, as in the 220 GHz -- 145 GHz difference map the
dust and IR emission is remarkably smaller, and the tSZ signal is
reduced since the non-relativistic tSZ brightness fluctuations vanish
at 218 GHz.

\begin{figure*}
\centering
\plotancho{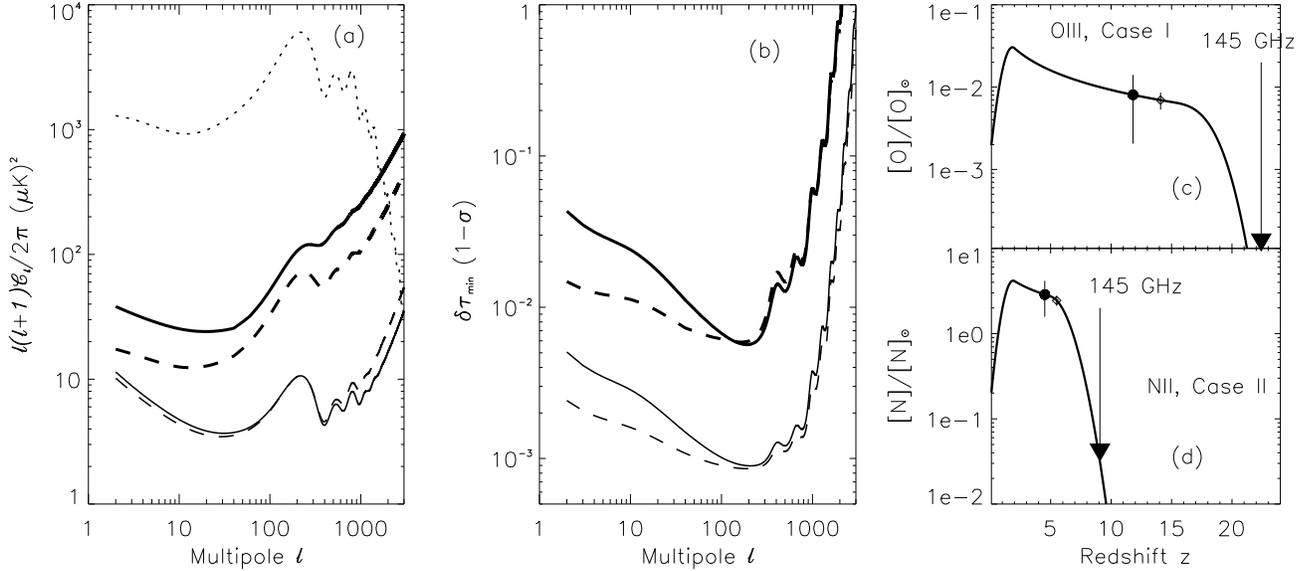}
\caption[fig:fig2]{ {\it (a)} Residual power spectra for the two map
 differences: 265 GHz - 145 GHz, (solid lines) and 220 GHz - 145 GHz
 (dashed lines). Thin lines correspond to the optimistic case, thick
 lines to the pessimistic case (see text for details).  The dotted
 line shows the reference CMB power spectrum. {\it (b)} Minimum
 detectable $\delta \teff^{mn}$ via a straight comparison of residual
 and CMB power spectra, i.e., $(\delta \teff^{mn})_{\ell} = |
 C_{\ell}^{res} / (2\; C_{\ell}^{CMB})|$, with $C_{\ell}^{res}$ the
 residual power as given in panel {\it (a)}. {\it c)} Constraints that
 ACT will be able to set if inter-channel calibration issues are not
 important. In a scenario with early ionization powered by massive Pop
 III UV emission, Oxygen will be twice ionized and ACT will be able to
 measure increments of its abundance of $\sim$ 3\% its solar value
 between redshifts 22 and 14, and increments of $\sim$ 12\% its solar
 value between redshifts 22 and 12. {\it (d)} In a late reionization
 scenario in which NII dominates the effective resonant scattering
 optical depth, ACT will detect increments in its abundance of $\sim$
 60\% its solar value between redshifts 9 and 5.5, and $\sim$ 2.3
 times its solar value between redshifts 9 and 4.5.

}
\label{fig:fig2}
\end{figure*}

\subsection{One step further: removing the tSZ}
\label{sec:tsz}

In Fig.~\ref{fig:fig1}c we see that the tSZ signal due to galaxy
groups and small clusters is the main contaminant. Since for most of
these sources the electronic temperature is such that relativistic
corrections are negligible, it should be possible to exploit the known
frequency dependence of the tSZ temperature fluctuations in order to
remove them from the difference map. This is feasible if the CMB
experiment has more than two frequencies, as it is the case for ACT.\\

For the sake of notation, let us refer to the 145 GHz, 220 GHz and 265
GHz channels of ACT as the $i$, $j$ and $k$ bands, respectively.  Let
$m$ and $n$ denote any arbitrary pair of ACT bands. The $i$ channel is
probing higher redshifts, and therefore will be used as the {\em
reference} channel, so metal-induced optical depths in the $j$ and $k$
channel can be written as $\teff^j = \teff^i + \delta \teff^{ji}$ and
$\teff^k = \teff^i + \delta \teff^{ki}$ respectively.  Given any pair
of ACT bands $m$ and $n$, it is possible to build the following
spatial template for tSZ signal:
\begin{equation}
\hat{\vs}_{tSZ}^{mn} = \frac{\vt_m-\vt_n}{\Delta g_{tSZ}^{mn}}.
\label{eq:tSZ_templ}  
\end{equation}
If the tSZ in band $n$ is $\vt^{tSZ}_n = g_{tSZ,n} \vs_{tSZ}$, then it
is clear that $\hat{\vs}_{tSZ}^{mn}$ will equal $\vs_{tSZ}$ only if
tSZ is the only component in the difference map, or dust and infrared
emission, together with the metal-induced signature, are negligible.
In such circumstances, the spatial power spectrum of this template
should be equal to that given in eq.(\ref{eq:deltaalmsq2}), and this
step should provide a useful consistency check. Next we consider the
difference map $\vt_j - \vt_i$. For this difference map, it is
possible to remove two different tSZ templates, namely
$\hat{\vs}_{tSZ}^{ki}$ and $\hat{\vs}_{tSZ}^{kj}$. That is:
\begin{eqnarray}
\corr^{ji}_{ki} & = & \vt^{j} - \vt^i - 
                    \Delta g_{tSZ}^{ji} \hat{\vs}_{tSZ}^{ki},\\
\corr^{ji}_{in} & = & \vt^{j} - \vt^{i} - 
                      \Delta g_{tSZ}^{ji} \hat{\vs}_{tSZ}^{kj}.
\label{eq:r1}
\end{eqnarray}
After introducing the dependencies on $\delta \teff^{ji}$ and $\delta
\teff^{ki}$, the last equations become
\begin{eqnarray}
\corr^{ji}_{kj} & = & -\vt^{CMB}\biggl[\delta \teff^{ji} 
   (1 - \frac{\Delta g_{tSZ}^{ji}}{\Delta g_{tSZ}^{kj}}) + 
               \delta \teff^{ki} \frac{\Delta g_{tSZ}^{ji}}{\Delta g_{tSZ}^{kj}}
                                     \biggr] + \delta \corr^{ji}_{kj},\\
\corr^{ji}_{ki} & = &  -\vt^{CMB}\biggl[\delta \teff^{ji} 
       (1 - \frac{\Delta g_{tSZ}^{ji}}{\Delta g_{tSZ}^{ki}}) + 
               \delta \teff^{ki} \frac{\Delta g_{tSZ}^{ji}}{\Delta g_{tSZ}^{ki}}
                                     \biggr] + \delta \corr^{mi}_{in}.
\label{eq:r2}
\end{eqnarray}
The equations for  $\corr^{ki}_{kj}$ and $\corr^{ki}_{ji}$ constitute an equivalent system of equations (i.e. they no not add more information).
The {\it errors} $\delta \corr^{ji}_{ki}$ and $\delta \corr^{ji}_{ji}$
contain residuals of the tSZ subtraction, together with linear
combinations of dust and IR galaxy signals. From our foreground model,
it is possible to estimate the covariance matrix of these
errors. Further, if we apply a template fit introduced in the last
part of Section (\ref{sec:fit}) on $\corr^{mi}_{mn}$ and
$\corr^{mi}_{in}$, we obtain estimates of the quantities
\begin{equation}
\beta^{ji}_{ki}  =  \biggl[\delta \teff^{ji} 
   (1 - \frac{\Delta g_{tSZ}^{ji}}{\Delta g_{tSZ}^{kj}}) + 
               \delta \teff^{ki} \frac{\Delta g_{tSZ}^{ji}}{\Delta g_{tSZ}^{kj}}
                                     \biggr],
\label{eq:r3a}
\end{equation}
\vspace*{.02cm}
\begin{equation}
\beta^{mi}_{in}  =  \biggl[\delta \teff^{ji} 
       (1 - \frac{\Delta g_{tSZ}^{ji}}{\Delta g_{tSZ}^{ki}}) + 
               \delta \teff^{ki} \frac{\Delta g_{tSZ}^{ji}}{\Delta g_{tSZ}^{ki}}
                                     \biggr].
\label{eq:r3b}
\end{equation}

Note these estimates for $\beta^{ji}_{ki}$ and $\beta^{mi}_{in}$ are
computed over the same area on the sky, and the correlation between
them arises due to the fact that foreground emission at different
frequencies is correlated. This is also the cause for the correlation
between $\delta\corr^{mi}_{ji}$ and $\delta \corr^{ji}_{ki}$, and for
this reason we shall approximate the correlation coefficient matrix of
$\beta^{ji}_{ki}$ and $\beta^{mi}_{in}$ by that of
$\delta\corr^{mi}_{ji}$ and $\delta \corr^{ji}_{ki}$.
 Finally, we must recall that we are not interested on
$\beta^{ji}_{ji}$ or $\beta^{ji}_{ki}$, but on $\delta \teff^{ji}$ and
$\delta \teff^{ki}$. The latter are just a linear transformations of
the former, and this makes straightforward the transformation of the
respective covariance matrices.  It is clear that this procedure will
be useful in case the tSZ component is far larger than the rest, since
it removes most of the tSZ present in the difference maps at the
expense of slightly increasing the level of the other contaminants.

\subsection{Sensitivity on $\delta \teff^{ji}$ and 
constraints on IGM enrichment}

ACT will scan a strip of 2.5$\degr$ width around the south celestial
pole, at a constant declination of $\delta \simeq -55.3 \degr$. The
resulting ring covering 400 square degrees will intersect the galaxy,
and hence the clean area will be smaller and located at high galactic
latitudes. In terms of dust emission, following the model of
\citet{fds}, we select the $\sim 140$ cleanest square degrees, and
group them into 83 patches of 500 pixels each. This will make us
sensitive to changes in the power spectrum in a multipole range $\ell
\in [80,2000]$.  We use the HEALPix\footnote{{\tt
http://healpix.jpg.nasa.gov}} \citep{healpix} pixelization at
resolution parameter $N_{side}=1024$. The linear pixel size is close
to 3.4 arcmins, and an effective convolution with a gaussian beam of
FWHM$=4$ arcmins is adopted for all components in all maps. We assume
a PSF reconstruction uncertainty ranging from 0.2\% (optimistic case)
to 1\% (pessimistic case) (i.e., $|\delta \epsilon^{mn}_{\ell m}| \sim
0.002 - 0.01 \;\forall l,m$) and consider how the cross-channel
calibration error affects the constraints.\\

When computing the CMB power spectrum we use the cosmological
parameters that best fit WMAP's first year data \citep{WMAP}, and as a
CMB template ($\vt^{CMB}$) we use one sky realization from such power
spectrum.  Foreground contamination is introduced according to
subsection \ref{sec:foreg}.  In what follows, in the pessimistic case
we assume PSF reconstruction error of 1\%, dust emission residuals of
30\% in the maps ($\sim 10$\% in $C_{\ell}$), 20\% error in IR sources
spectra index, no tSZ subtraction (after masking).  In the optimistic
case we assume $|\delta \epsilon_{\ell m}^{mn}| \sim 2\times 10^{-3}$,
10\% error in dust amplitude modeling, $\delta \alpha_{BLAST} /
\alpha_{BLAST}$, of 5\%, tSZ subtraction down to a 10\% in the
maps. \\


In Fig.(\ref{fig:fig2}a) we show the power spectra of the foreground
residuals, for each of the two map differences, $ji$ (or 265 - 145
GHz, solid lines) and $ki$ (220 - 145 GHz, dashed lines).
%
Thick lines correspond to the pessimistic case, thin lines to the
optimistic case. We can see that the residuals power spectra become
proportional to the CMB power spectra at $l\sim 200$, and that this
fact is more visible when foreground subtraction is more
accurate. This is due to the fact that PSF reconstruction errors
$\delta \epsilon^{ji}_{\ell m}$ are taken to have, on average, the
same amplitude for all multipoles $\ell$'s. Hence $|\delta
\epsilon^{ji}_{\ell m} a_{\ell m}^{CMB}|^2 \propto C_{\ell}^{CMB}$.\\

We forecast at what precision the $\delta \teff^{mn}$ can be recovered
via two different approaches. The first approach applies the template
fit outlined in Section \ref{sec:fit} on the residual level given by
our pessimistic of foregrounds, whereas the second provides the
sensitivity on $\delta \teff^{mn}$ when the template fit is applied on
the optimistic case. The errors on $\delta \teff^{mn}$ are shown in
Table~\ref{tab:tab1} for the two map differences considered: $ji$ (220
- 145 GHz) and $ki$ (265 - 145 GHz). Note that the ratio $\delta
C_{\ell}^{mn} / (2C_{\ell}^{CMB})$ can also provide a (less accurate)
guess on $\delta \teff^{mn}$. Finally, let us remark that these three
ways of estimating the increment of the metal induced optical depth,
although not independent, can provide useful mutual consistency
checks. For experiments scanning larger areas (i.e., SPT), the errors
in the template fit method would roughly scale like the inverse square
root of the area covered.\\

\begin{table}
\caption{Sensitivity on $\delta \teff^{mn}$}
\begin{center}
\begin{tabular}{ccc}
\hline
\hline
&   Pessimistic case & Optimistic case    \\
\hline
$\sigma_{\delta \teff^{ji}}$   &  $1.9 \times 10^{-3}$  &  $2.2\times 10^{-4}$   \\
\hline
$\sigma_{\delta \teff^{ki}}$   &  $2.7 \times 10^{-3}$ &  $3.0\times  10^{-4}$  \\
\hline\hline
\end{tabular}
\end{center}
\label{tab:tab1}
\end{table}

 The errors in Table~\ref{tab:tab1} are given assuming that
  cross-channel calibration uncertainties are negligible. Thus if
  cross-channel calibration uncertainties ($\delta f_{ji}$, $\delta
  f_{ki}$) are worse than the values quoted in Table ~\ref{tab:tab1},
  then they will become the limiting factors in the measurements, (as
  already shown in eq.(\ref{eq:deltaalm2})). Since abundances are
  proportional to optical depths (see eq.(\ref{eq:tausob1})), it
  follows that errors in abundances are proportional to cross-channel
  calibration errors in case the latter dominate over all other
  sources of errors. These levels are reachable with current Fourier
  Spectrometer technology, \citep{fspec1,fspec2}.\\

%
If indeed inter-channel calibration is not an issue, then by comparing
Table~\ref{tab:tab1} with Table\~1 in BHMS, we can infer that ACT will
be able to set interesting constraints on the enrichment history. If
Pop III stars ionized Oxygen twice, ACT will be able to detect
increments of OIII abundances of $\sim$ 3\% its solar value between
$z\sim 22$ and $z\sim 14$, and of $\sim 12\%$ its solar value between
$z\sim 22$ and $z\sim 12$, at 95\% confidence level, (see
Fig.~\ref{fig:fig2}c). Note that these redshift ranges are
particularly interesting, since they are close to the upper range for
reionization redshift as indicated by {\it WMAP} ($z_{\rm re}\simeq 11
\pm 2.5$ or $5<z_{\rm re}<15$ at 2-$\sigma$ level).  Further, if
reionization happens at the lower end of the WMAP values, and Nitrogen
dominates the effective optical depth to CMB photons, then ACT will
set constraints on the abundance of NII, since it can measure
increments of $\sim$ 60\% its solar value between $z\sim 9$ and $z\sim
5.5$, and of $\sim$ 2.3 times its solar value between $z\sim 9$ and
$z\sim 4.5$, (95 \% c.l., see Fig.~\ref{fig:fig2}d).

\section{Discussion and Conclusions}

The probability that a CMB photons is scattered by a given metal
species depends primarily on two things: how abundant such species is
and how likely the transition is. Indeed, as already shown in
eq.(\ref{eq:tausob1}), the Sobolev optical depth for the resonant
transitions is proportional to the number density of ($n_i(z)$) the
intervening species and the so-called oscillation strength of the
transition ($f_i$).
This means that for fixed metal abundance, the oscillator strength,
the transition wavelenght, and the relative population of the lower
level involved in the transition will rule which lines are more {\it
visible}.  From Table ~ 1 in BHMS, the transitions giving rise to
higher optical depth correspond to OIII (88.36 $\mu$m, 51.81 $\mu$m),
OI (63.18 $\mu$m), CII (157.74 $\mu$m) and NII (205.30 $\mu$m).  For
the frequency range of ground based experiments, O and N are the
elements more likely to generate most of the optical depth to CMB
photons.  (Note that, although CII shows a strong transition, current
Pop-III yield models predict low C/N abundance ratios,
\citep{Meynet}).\\

In the presence of only one species, say O, the technique presented
here measures the increase of O abundance between some higher redshift
bin (corresponding to the lowest frequency of the experiment) and
lower redshift bins (corresponding to lower frequency bands).  Under
the additional assumption that the metal abundance at the highest
redshift bin is negligible, the method yields constraint on the
species abundance as a function redshift.  In presence of different
species (say O and N), this technique will be sensitive to the
integrated effect of the different species. Note however that the same
frequency band probes different redshift for resonant transitions of
different elements.  For example for OIII 145 GHZ corresponds to $z
\simeq 22$, 220 GHz to $z \simeq 14$ and 265 GHz to $z \simeq 12$, but
for NII, 145 GHZ corresponds to $z \simeq 9$, 220 GHz to $z \simeq
5.5$ and 265 GHz to $z \simeq 4.5$. \\

To interpret these measurements in light of the reionization process,
we need a model of metal production for pop-III stars and an estimate
of star formation rates at these redshifts. There have been recent
advances in the computation of the yields for pop III stars
\citep{Meynet}: a more careful modeling including the effect of
rotation in the evolution of massive, low metallicity stars makes
predictions for the properties of low metallicity halo stars in better
agreement with observations.  Star formation rate estimates are
somehow dependent on the initial mass function, however the yields are
relatively insensitive \citep{Meynet} to the mass of the star for
massive ($M>10 M_{\odot}$) stars and pop-III are expected to be
massive \citep{Abel}. Therefore a constraint on the metal abundance
obtained from measurements of the optical depth for resonant
scattering, will translate into a constraint on the number of pop-III
stars. Different reionization scenarios produce a different amount of
pop-III stars as a function of redshift. While a detailed calculation
along these lines will be presented elsewhere, we can show the
potential impact of this technique and its complementarity with other
probes of reionization with the following example. The nature of
ionizing sources is not well understood
(e.g. \citet{BarkanaLoeb2001}).  A detection of $\sim$ 12\% solar
metallicity of oxygen at $z\sim 12$ would imply that pop-III stars
played a major role in reionization and thus rule out the scenario of
reionization by mini quasars. \\

In addition, the method presented here of tracing the reionization
history with a measurement if the IGM enrichment, is complementary to
other probes of reionization. For example the CMB is sensitive to
ionizing photons produced during reionization and their escape
fraction, and the horizon scale at reionization redshift.  H 21 cm
radiation probes neutral Hydrogen and the Ly-$\alpha$ forest has provided
measurements of metallicities at $z \sim 4 - 5·$.  All these methods rely on
different physics and thus provide a invaluable tool for understanding
the physics of reionization. In addition they are affected by
foregrounds and systematics of very different nature, so their
comparison provides a useful consistency check for the results.\\

%

The forecasts presented here are for an experiment designed with
another goal in mind. It is interesting to ask how well an ideal
experiment or a dedicated experiment could perform.  A dedicated
experiment would observe through the same PSF at different
frequencies, in order to minimize the effects of PSF reconstruction
errors.  An self-calibrator device would assure accurate inter-channel
calibration.  The set of observing frequencies should be a compromise
between accurate tSZ subtraction (this forces the ratios
$g_{tSZ,i}/g_{tSZ,j}$, $g_{tSZ,i}/g_{tSZ,k}$ to be {\it far} from
unity) and the choice of an interesting redshift range in terms of
cosmic enrichment history (which is determined by transition
frequencies of the metal species dominating the resonant scattering).
Detector spectral widths verifying $\Delta \nu / \nu\; ^{>}_{\sim}$
0.1 should assure that resonant scattering is dominant over
collisional emission (see discussion in BHMS).\\

In our optimistic model for foregrounds, if the PSF reconstruction
  errors are smaller or equal than $5\times 10^{-4}$, then foreground
  residuals are the limiting factors and the constraints on the
  abundances improve by a factor of three compared to the case
  considered here, for the same sky coverage.  This sets the target
  for PSF reconstruction for a dedicated experiment. In this case the
  cross-cannel calibration requirements also becomes stricter by a
  factor four, achievable with current Fourier Spectrometer technology.
  Consider OIII and NII: the transitions quoted above (88.3 $\mu$m and
  205.30 $\mu$m, respectively) will dominate the effective optical
  depth to CMB photons because of the pop III yields for these
  elements and because of the transition oscillator strength.  For
  these elements, there is an optimal frequency range to probe
  reionization. The upper limit to the redshift of reionization sets
  the reference channel to be at $\gap 100$GHz.  On the other end, at,
  say, 350 GHz, the level of contamination due to dust in the Galaxy
  and IR sources becomes important. We conclude that the optimal
  frequency window will be between 100 GHz and 350 GHz. This range
  fits well with the frequency coverage of Planck, and, at the same
  time, is well accessible by ground-based experiments.\\



In summary, we have presented an algorithm to realistically measure
the signature of metals in the power spectrum of the CMB. We have
shown how the different foregrounds can be removed and have presented
a method to combine the maps at different frequencies optimally. The
dominant foreground is the tSZ and the dominant source of error is the
beam error and cross-channel calibration. We have forecasted the
performance of the algorithm for an experiment with characteristics
similar to ACT and have shown that one can constrain IGM abundances
larger than 1\% solar at $z > 10$. Since the method probes the metal
abundance, and therefore the star formation rate, at the redshift
corresponding to the frequency of the band, it is possible to do
tomography of the metallicity enrichment process.  Multifrequency CMB
experiments can provide a tomographic reconstruction of the
re-ionization epoch that is nicely complementary to other reionization
probes.

\section*{Acknowledgments}

We thank Mark Devlin and Lyman Page for useful discussions on
instrumentation and systematics control and Rashid A. Sunyaev for
discussions that motivated this work and comments and criticism. Some
of the results in this paper have been derived using the
HEALPix\footnote{HEALPix URL site: {\it http://healpix.jpg.nasa.gov}}
\citep{healpix} package.  We acknowledge the use of the Legacy Archive
for Microwave Background Data Analysis (LAMBDA). Support for LAMBDA is
provided by the NASA Office of Space Science. This research is
supported in part by grant NSF AST-0408698 to the Atacama Cosmology
Telescope.  CHM and LV are supported by NASA grants ADP03-0000-0092
and ADP04-0000-0093. The work of RJ is supported by NSF grants
AST0206031, AST-0408698, PIRE-0507768 and NASA grant NNG05GG01G.

\label{lastpage}




\end{document}